\documentclass[reprint,amssymb,prx,longbibliography]{revtex4-2}
\usepackage[utf8]{inputenc}
\usepackage[T1]{fontenc}
\usepackage{graphicx}
\usepackage{amsmath}
\usepackage{amssymb}
\usepackage{mathrsfs}
\usepackage{braket}
\usepackage{geometry}
\usepackage{here}
\usepackage{lmodern}
\usepackage{array}
\usepackage{natbib}
\usepackage{url}
\usepackage{textcase}
\usepackage{bm}
\usepackage{natbib}
\usepackage[usenames,dvipsnames]{color}
\begin{document}
\title{Generalized multichannel optical theorem: Coherent control of the total scattering cross section}
\author{Adrien Devolder$^{1}$, {Timur V. Tscherbul$^{2}$}, and Paul Brumer$^{1}$}

\affiliation{$^{1}$Chemical Physics Theory Group, Department of Chemistry, and Center for Quantum Information and Quantum Control, University of Toronto, Toronto, Ontario, M5S 3H6, Canada\\
	$^{2}$Department of Physics, University of Nevada, Reno, NV, 89557, USA}
\begin{abstract}
The optical theorem is a fundamental aspect of quantum scattering theory. Here, we generalize this theorem to the case where the incident scattering state is a superposition of internal states of the collision partners, introducing additional interference contributions and, e.g.,  providing a route to control the total integral cross section. As in its standard form, forward scattering plays an essential role in the generalized optical theorem, but with interference terms being related to the inelastic forward scattering amplitudes \emph{between} states in the initial superposition. Using the resultant control index, we show that extensive control is possible over ultracold collisions of oxygen molecules in their rovibrational ground states, and of $^{85}$Rb-$^{85}$Rb collisions, promising systems for the first experimental demonstration of the quantum interference control of the total scattering cross section. 
\end{abstract}

\date{\today}
	\maketitle
	\flushbottom
\section{Introduction} 
The multichannel optical theorem is a remarkable result in scattering theory that relates the imaginary part of the elastic forward amplitude to the total scattering cross section \cite{Book_Taylor,Book_Friedrich}, a consequence of probability conservation during the scattering. No matter how complicated the scattering, all information about the total cross section is contained in the elastic forward scattering. Discovered first for light by Sellmeier \cite{Sellmeier1871} and Rayleigh \cite{Strutt1871}, the optical theorem was subsequently extended to quantum mechanics by Feenberg \cite{Feenberg1932} and Bohr et al.\cite{Bohr1939}{, and has been used in many areas, such as atomic, molecular and optical physics \cite{Joshipura1996,Jiang1995}, plasma physics \cite{Scholz1990}, astrophysics \cite{Jiang1995}, atmospheric physics \cite{Joshipura1996,Jiang1995}, nuclear physics\cite{Kim1997,Lipperheide1987,Udagawa1981} and high-energy physics \cite{Pancheri2016}}.

The standard optical theorem assumes an initial pure state of fixed energy. Here we extend this theorem to a broader class of initial states, a superposition at a fixed energy that generates new and interesting interference contributions. We then demonstrate the utility of the generalized theorem in controlling the total scattering cross section. Examples of efficient control of ultracold atomic and molecular collisions are provided. 

Over the past decades, progress in cooling techniques has enabled the creation of ultracold gases of atoms and molecules\cite{Book_Krems_2}. Nevertheless, loss of coherence is caused by collisions between the atoms/molecules  {and limits their use in quantum information science \cite{Karra2016,Park2017,Yu2019,Albert2020}}. On the other hand, such collisions are essential for understanding ultracold chemistry \cite{Carr2009,Heazlewood2021}. Fortunately, some ultracold collisions can be controlled due to the ability to fully define the internal states of the atoms/molecules, and due to the small value of the kinetic energy relative to the perturbations induced by external fields. As a consequence, the vast majority of ultracold control scenarios are based on external fields (magnetic, electric or optical) \cite{Chin2010,Book_Krems}, with molecules/atoms prepared in a well-defined internal state. In these strategies, the control knobs are the field parameters (strength, frequency, etc.). 

However, this approach, as noted below, has deficiencies, motivating a different strategy. That is, the control of the internal degree of freedom at ultracold temperature enables the preparation of quantum superpositions of internal states that can be used to induce interferences between scattering amplitudes, in analogy with the double slit experiments \cite{Brumer_book,Zhou2021}. 											    
 Then, instead of control via variation of field parameters, ultracold collisions are controlled by changing the nature of the initial superposition. The resultant effect on the system is the principle of coherent control \cite{Brumer_book}.  
 
 In previous work, we demonstrated that the ultracold regime is ideal for coherent control and that control can be achieved for resonant processes such as collisional spin exchange \cite{Devolder:21}, which can be completely suppressed (or activated), via destructive (constructive) interference. Control is achieved without persistent application of external perturbations. In particular, the collision partners (atoms or molecules) need not have electric and(/or) magnetic dipole moments to be coherently controlled. For example, collisions of nonmagnetic homonuclear molecules, like H$_2$ or Sr$_2$, could be manipulated. Moreover, the absence of external field could be important for high-precision measurements (for example with Sr$_2$ molecules \cite{Zelevinsky2008,Kotochigova2009}) where suppressing external perturbations is significant.  

At present, contrary to the unimolecular processes \cite{Zhu1995}, the observation of coherent control of bimolecular processes is still an open experimental challenge due to a number of issues. First, the preparation of the initial superposition is experimentally challenging. Conditions on the coherent control of scattering events require either entanglement of the external and internal degrees of freedom \cite{Shapiro1996,Abrashkevich2001,Gong2003,Zeman2004}, or superposition of degenerate magnetic sublevels (an $m$-superposition) \cite{Brumer1999}. However, recently, some progress has been made in preparing $m$-superpositions in H$_2$ molecules and its isotopologues \cite{Perreault:17,Perreault:18,Mukherjee2011}. Secondly, our previous study of coherent control at ultralow temperatures \cite{Devolder:21} focused on state-to-state cross sections, and indicated a need for complicated coincidence measurements of the two scattered molecules \cite{Liu2021,Margulis2020}. On the other hand, the total cross section, considered here, could be experimentally easier to measure and control. 

Below we demonstrate the utility of the generalized theorem in controlling the total scattering cross section. The theorem allows to answer the following fundamental questions: How does the standard optical theorem generalize when the scattering is of an initial superposition of internal states? What are the new insights for coherent control of the total cross section arising from this generalization? And, with these new insights, can we identify atomic and/or molecular collisions promising for a first experimental demonstration of coherent control of the total scattering cross section? Answering these questions is important for the development of this completely new control strategy for ultracold collisions and for applications in other scattering scenarios. 

The structure of the paper is as follows. We first derive the generalized multichannel optical theorem (Section II) and, in the rest of the paper, use the resultant theorem to address the control related issues raised above. In Section III, we compare the derived formulas with the standard form of the optical theorem. The generalized optical theorem allows us to define a coherent control index for the total integral cross section (ICS) in Section IV, which we then use to analyze promising systems for experiments in Section V. Exact scattering calculations are presented which show impressive control over the total cross section for the experimentally realizable ultracold $^{85}$Rb-$^{85}$Rb and $^{17}$O$_2$-$^{17}$O$_2$ collisions. We conclude in section VI. 

\section{Derivation}
Consider an initial superposition of $N_{sup}$ degenerate internal states of the scattering partners denoted $\ket{i}=\ket{\nu_A}\ket{\nu_B}$. Here $\nu_{A,B}$ are the quantum numbers characterizing the internal states of the molecular or atomic collision partners A+B. The initial state is:
\begin{equation}
\Psi_{in}(\vec{r},\xi)=\textrm{e}^{ikz} \sum_{i=1}^{N_{sup}} a_i \ket{i},
\label{eq:psi_ini}
\end{equation}
where the $z$-axis is defined along the initial relative momentum $\vec{k}$. Here $a_i$ are the superposition coefficients, $\vec{r}=(r,\theta,\phi)$ is the relative position between the two collision partners, and $\xi$ is composed of all internal coordinates contained in $\ket{i}$.

After the collision, the system is in a superposition of scattered spherical waves in various open channels \cite{Book_Taylor}:
\begin{equation}
	\Psi_{out}(\vec{r},\xi)=\sum_{j} f_{sup \rightarrow j}(\theta,\phi) \frac{\textrm{e}^{ik_jr}}{r}\ket{j},
\label{eq: psi_after}
\end{equation}
where $k_j$ is the final relative momentum in the state $\ket{j}$ and 
\begin{equation}
f_{sup\rightarrow j}(\theta,\phi)=\sum_{i=1}^{N_{sup}}a_if_{i\rightarrow j}(\theta,\phi)
\label{eq:scatt_amp_sup}
\end{equation} is the scattering amplitude from the initial superposition to the final state $\ket{j}$. Here, $f_{i\rightarrow j}$ is the scattering amplitude from state $\ket{i}$ to $\ket{j}$.
The overall wavefunction therefore obeys the boundary condition:
\begin{equation}
\Psi(\vec{r},\xi) \xrightarrow[r\to \infty]{} \textrm{e}^{ikz} \sum_{i=1}^{N_{sup}} a_i \ket{i}+\sum_{j} f_{sup \rightarrow j}(\theta,\phi) \frac{\textrm{e}^{ik_jr}}{r}\ket{j}.
\label{eq:bound_cond}
\end{equation}

The optical theorem can be derived by imposing conservation of probability via the continuity equation:
\begin{equation}
\oint \vec{j}(\vec{r})\cdot\hat{e}_r r^2d\Omega =0,
\end{equation}
where $\vec{j}(\vec{r})=\int \textrm{d}\xi\frac{\hbar}{2i\mu}\Psi^*(\vec{r},\xi)\vec{\nabla}\Psi(\vec{r},\xi)+\textrm{c.c}$ is the current density, $\mu$ is the reduced mass, $\hat{e}_r$ is the unit radial vector, $d\Omega=\,\sin\theta \textrm{d}\theta \textrm{d}\phi$ is the solid angle, and c.c denotes complex conjugate. The continuity equation states that the scattering flux through any closed surface must vanish, which is also the case for a sphere with radius $r\rightarrow \infty$ where the boundary condition (\ref{eq:bound_cond}) is imposed. This flux, defined as
$I\equiv \lim_{r\rightarrow\infty}\oint \vec{j}(\vec{r}).\hat{e}_r r^2\textrm{d}\Omega $, can be expanded into three terms: the incoming contribution $I_{in}$, the outgoing contribution $I_{out}$ and the interference contribution $I_{int}$:
\begin{equation}
I_{in}+I_{out}+I_{int}=0.
\label{eq:cons}
\end{equation}
The contributions $I_{ini}$, $I_{out}$ and $I_{int}$ are derived in the appendices A, B and C, respectively, and their final values are: 
\begin{equation}
I_{in}=0,
\label{eq:I_in}
\end{equation}
\begin{equation}
I_{out}=\frac{\hbar k}{\mu}\sigma_{sup}^{tot},
\label{eq:I_out}
\end{equation}
\begin{equation}
I_{int}=\frac{-4\pi\hbar}{\mu} \sum_{i,i'=1}^{N_{sup}} \textrm{Im}[a_i^*a_{i'}f_{i' \rightarrow i}(\theta=0)].
\label{eq:I_int}
\end{equation}

Introducing (\ref{eq:I_in}),(\ref{eq:I_out}) and (\ref{eq:I_int}) in Eq. (\ref{eq:cons}), one obtains the total integral cross section (ICS):
\begin{equation}
\sigma_{sup}^{tot}=\frac{4\pi}{k}\sum_{i,i'=1}^{N_{sup}} \textrm{Im}[a_i^*a_{i'}f_{i' \rightarrow i}(\theta=0)].
\label{eq:optical_theorem1}
\end{equation}
Consider then the dependence of the total cross section on the relative phases $\beta_{i}$ between the states of the initial superposition, which can be explicitly illustrated by writing the superposition coefficients in their polar form: $a_i=|a_i|\textrm{e}^{i\beta_i}$:
\begin{equation}
\begin{split}
\sigma_{sup}^{tot}=&\frac{4\pi}{k}\sum_{i=1}^{N_{sup}}|a_i|^2\textrm{Im}[f_{i \rightarrow i}(\theta=0)]\\+&\frac{4\pi}{k}\sum_i\sum_{i'\neq i}|a_i||a_{i'}| \textrm{Im}\left[\textrm{e}^{i(\beta_{i'}-\beta_{i})}f_{i' \rightarrow i}(\theta=0)\right].
\label{eq:optical_theorem3}
\end{split}
\end{equation}
Finally, the symmetric relation between the scattering amplitudes $f_{i' \rightarrow i}=f_{i \rightarrow i'}$ can be exploited to obtain the generalized optical theorem:
\begin{equation}
\begin{split}
\sigma_{sup}^{tot}=&\frac{4\pi}{k} \sum_{i=1}^{N_{sup}}|a_i|^2 \textrm{Im}[f_{i \rightarrow i}(\theta=0)]\\+&\frac{8\pi}{k} \sum_{i=1}^{N_{sup}}\sum_{i'>i}|a_i| |a_{i'}|\textrm{Im}[f_{i \rightarrow i'}(\theta=0)] \,\cos(\beta_{i'}-\beta_{i}).
\label{eq:optical_theoremsym}
\end{split}
\end{equation}
The symmetric form is valid when the time-reversal symmetry applies. When it does not, Eq. (\ref{eq:optical_theorem3}) should be used.

Equations (\ref{eq:optical_theorem3}) and (\ref{eq:optical_theoremsym}) are the central result of this work. They establish a relation between the magnitude of the total ICS and the preparation coefficients of the internal superposition, enabling coherent control of the total ICS. Note that they are of the standard coherent control form \cite{Brumer_book}, indirect scattering terms plus interference between pairs of states.

\section{A generalization of the standard optical theorem}

The generalized optical theorem (Eq. \ref{eq:optical_theoremsym}) establishes the total ICS as composed of an incoherent contribution and a coherent contribution:
\begin{equation}
\sigma_{incoh}=\frac{4\pi}{k} \sum_{i=1}^{N_{sup}}|a_i|^2 \textrm{Im}[f_{i \rightarrow i}(\theta=0)],
\label{eq:incoherent}
\end{equation}
\begin{equation}
\sigma_{coh}=\frac{8\pi}{k} \sum_{i=1}^{N_{sup}}\sum_{i'>i}|a_i| |a_{i'}|\textrm{Im}[f_{i \rightarrow i'}(\theta=0)] \,\cos(\beta_{i'}-\beta_{i}).
\label{eq:int}
\end{equation}
The incoherent contribution is related to the elastic forward scattering amplitudes of individual states in the superposition, weighted by their populations. This contribution would be the same for a classical mixture with $|a_i|$ given, for example by Boltzmann populations. The standard optical theorem without superposition follows from the incoherent contribution (\ref{eq:incoherent}) in the limit $N_{sup}=1$, $a_1=1$:
\begin{equation}
\sigma_1^{tot}=\frac{4\pi}{k}\textrm{Im}[f_{1 \rightarrow 1}(\theta=0)].
\end{equation}
 
 For the purpose of control, the significance of the coherent contribution (\ref{eq:int}) is that it allows control over the total ICS by varying the amplitude product $|a_i| |a_{i'}|$ and the relative phases $(\beta_{i'}-\beta_{i})$  between the states in the initial superposition. The coherent contribution (\ref{eq:int}) is related to the inelastic forward scattering amplitudes \emph{between} the states in the initial superposition, and is the primary attribute of the generalized optical theorem, corresponding to interference between scattering events in different channels. \emph{It indicates that inelastic scattering between the channels involved in the initial superposition is a prerequisite for coherent control of total ICS.} Note that whatever the system, the coherent terms oscillate as $\cos(\beta_{i'}-\beta_{i})$ and, that contrary to the elastic component $\textrm{Im}[f_{i \rightarrow i}(\theta=0)]$, the inelastic component $\textrm{Im}[f_{i \rightarrow i'}(\theta=0)]$ can be negative.
 
Consider now the key quantity that determine the magnitude of the coherent contribution to the total ICS, (\ref{eq:int}), the imaginary part of the inelastic forward scattering amplitude:
 \begin{equation}
\textrm{Im}[f_{i \rightarrow i'}(\theta=0)]=\frac{1}{4k}\sum_j\sum_{\ell',m'}  \tilde{T}_{i\rightarrow j,\ell',m'}\tilde{T}_{i'\rightarrow j,\ell',m'}^*,
\label{eq:sum_T_Im_inelas}
\end{equation}
where $\tilde{T}_{i\rightarrow j,\ell',m'}=\sum_{\ell}i^\ell \sqrt{2\ell+1}T_{i,\ell,0\rightarrow j,\ell',m'}$, $\ell$ and $\ell'$ are the initial and final partial wave, and $T_{i,\ell,0\rightarrow j,\ell',m'}$ are the $T$-matrix elements. Equation (\ref{eq:sum_T_Im_inelas}) provides an important perspective, that is: \emph{interference is a result of scattering of states $\ket{i}$ and $\ket{i'}$ into the same states $j,\ell',m'$}. Further, the sum $ \sum_j\sum_{\ell',m'}  \tilde{T}_{i\rightarrow j,\ell',m'}\tilde{T}_{i'\rightarrow j,\ell',m'}^*$ is seen to be real. Equation (\ref{eq:sum_T_Im_inelas}) is required by conservation of probability and the generalized optical theorem can also be proven via this relation.
 \section{Coherent control of the total cross section}
\subsection{Two-state superpositions}
 The generalized optical theorem allows for considerable new insights into the coherent control of the total ICS. First, consider the case where the initial superposition of the scattering partners is composed of two states, $\Psi_{in}(\vec{r},\xi)=\textrm{e}^{ikz} \left(\cos\eta\ket{1}+\sin\eta \textrm{e}^{i\beta}\ket{2}\right)$. Here, control is achieved by changing the amplitude and phase of the superposition by varying $\eta$ and $\beta$, respectively. The generalized optical theorem [Eq. (\ref{eq:optical_theoremsym})] takes the form:
 \begin{equation}
 \begin{split}
 \sigma_{sup}^{tot}=&\frac{4\pi}{k}\{\cos^2(\eta) \textrm{Im}[f_{1 \rightarrow 1}]+\,\sin^2(\eta) \textrm{Im}[f_{2 \rightarrow 2}]\}\\+&\frac{8\pi}{k} \,\cos\eta \,\sin\eta \ \textrm{Im}[f_{1 \rightarrow 2}] \,\cos\beta.
 \end{split}
 \label{eq:optical_theorem_2st}
 \end{equation} 
 Insight is afforded by minimization and maximization of this expression. The optimization with respect to $\beta$ is straightforward; maximal for $\cos\beta=1$ ($\beta_{max}=0$) and minimal for $\cos\beta=-1$ ($\beta_{max}=\pi$) if $\textrm{Im}[f_{1 \rightarrow 2}(\theta=0)]$ is positive. It is the opposite if $\textrm{Im}[f_{1 \rightarrow 2}(\theta=0)]$ is negative. The optimization with respect to $\eta$ gives:
 \begin{equation}
 \eta_{min}=\frac{1}{2} \arctan\left(\frac{2 \left|\textrm{Im}[f_{1 \rightarrow 2}]\right|}{\textrm{Im}[f_{2 \rightarrow 2}]-\textrm{Im}[f_{1 \rightarrow 1}]}\right),
 \label{eta_min}
 \end{equation}
   \begin{equation}
 \eta_{max}=-\frac{1}{2} \arctan\left(\frac{2 \left|\textrm{Im}[f_{1 \rightarrow 2}]\right|}{\textrm{Im}[f_{2 \rightarrow 2}]-\textrm{Im}[f_{1 \rightarrow 1}]}\right),
 \label{eta_max}
 \end{equation} 
 Introducing $\eta_{min}$,$\beta_{min}$, $\eta_{max}$ and $\beta_{max}$ in the Eq. (\ref{eq:optical_theorem_2st}), one obtains the minimal and maximal values of the total ICS:
 \begin{equation}
 \begin{split}
 \sigma_{min}^{tot}=&\frac{2\pi}{k}\left(\textrm{Im}[f_{1 \rightarrow 1}]+\textrm{Im}[f_{2 \rightarrow 2}]\right)\\-&\frac{2\pi}{k}\sqrt{\left(\textrm{Im}[f_{2 \rightarrow 2}]-\textrm{Im}[f_{1 \rightarrow 1}]\right)^2+4 \textrm{Im}^2[f_{1 \rightarrow 2}]}
 \end{split},
 \end{equation}

 \begin{equation}
 \begin{split}
 \sigma_{max}^{tot}=&\frac{2\pi}{k}\left(\textrm{Im}[f_{1 \rightarrow 1}]+\textrm{Im}[f_{2 \rightarrow 2}]\right)\\+&\frac{2\pi}{k}\sqrt{\left(\textrm{Im}[f_{2 \rightarrow 2}]-\textrm{Im}[f_{1 \rightarrow 1}]\right)^2+4 \textrm{Im}^2[f_{1 \rightarrow 2}]}
 \end{split}.
 \end{equation}
The extent of the control is determined by the magnitude of $\textrm{Im}[f_{1 \rightarrow 2}]$, a quantity bounded between zero and $\sqrt{\textrm{Im}[f_{1 \rightarrow 1}]\textrm{Im}[f_{2 \rightarrow 2}]}$. That leads us to define a control index:
\begin{equation}
R_c=\frac{|\textrm{Im}[f_{1 \rightarrow 2}]|}{\sqrt{\textrm{Im}[f_{1 \rightarrow 1}]\textrm{Im}[f_{2 \rightarrow 2}]}},
\label{eq:Nc_2}
\end{equation}
which ranges from zero to one. When $R_c=1$, the Schwartz equality ($\left|\textrm{Im}[f_{1 \rightarrow 2}]\right|=\sqrt{\textrm{Im}[f_{1 \rightarrow 1}]\textrm{Im}[f_{2 \rightarrow 2}]}$) holds. In this case, the minimum value of the total cross section $\sigma_{min}^{tot}$ vanishes while the maximal value is the sum of the total cross sections in absence of superposition, $ \sigma_{max}^{tot}=\frac{4\pi}{k}\left(\textrm{Im}[f_{1 \rightarrow 1}]+\textrm{Im}[f_{2 \rightarrow 2}]\right)=\sigma_1^{tot}+\sigma_2^{tot}$. Hence, the value of $R_c$ allows us to quantify the extent of coherent control of the total ICS and interpret when systems display the maximum possible degree of control (which is realized for $R_c$=1). An example of complete control is the case of ideal resonance, \cite{Zeman2004}, where the resonance occurs for all final states. Another favorable situation is when the number of channels significantly populated in a collision is equal to (or less than) the number of states in the superposition\cite{Frishman1999a}. The latter case is illustrated below in the section V. 
\subsection{$N_{sup}$-state superpositions ($N_{sup}$>2)}
   \begin{figure*}
	\centering
	\includegraphics[width=1.4\columnwidth]{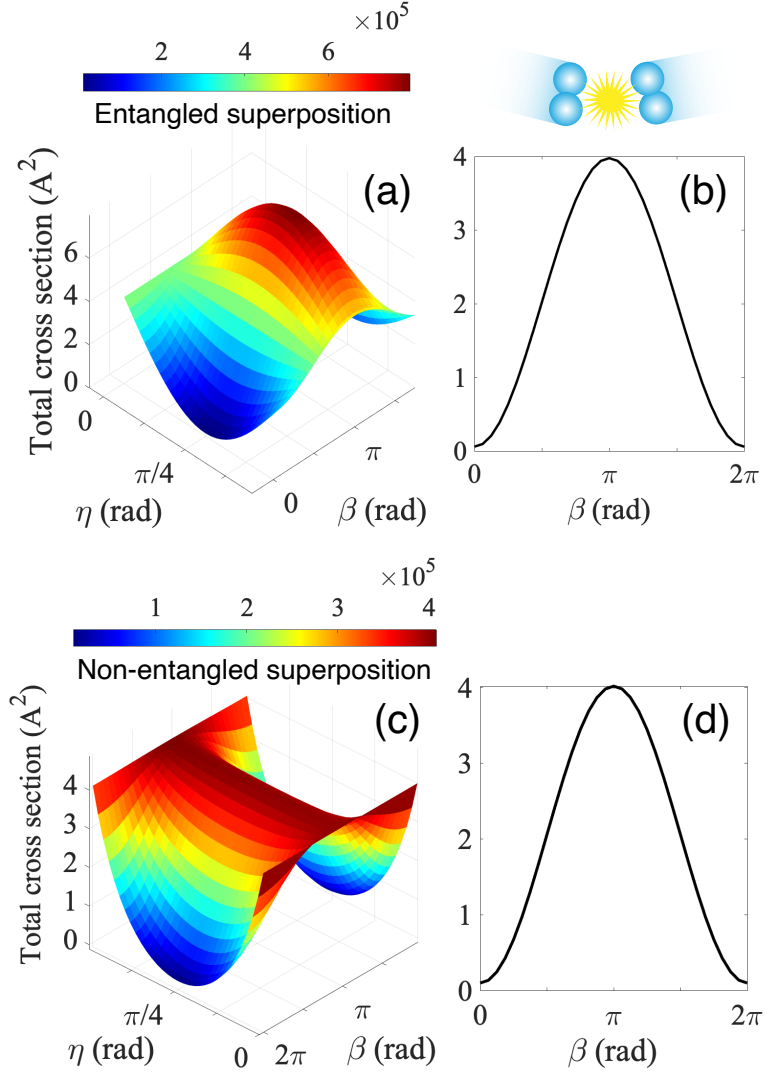}
	\caption{Coherent control of the total integral cross section for the cold $^{17}$O$_2$+ $^{17}$O$_2$ collisions at 10 $\mu$K from the initial superposition $\Psi_{E}$ (upper panels) and $\Psi_{NE}$ (lower panels). (a) and (c): Control landscape; (b) and (d): Control by the relative phase $\beta$ with $\eta=\pi/4$ fixed.}
	\label{fig:O2O2}
\end{figure*}
The optimization can be generalized to a superposition of $N_{sup}$ states via a procedure similar to that introduced in Ref. \cite{Frishman1999a}. We define the matrix $\Sigma_{ij}=\frac{4\pi}{k}\textrm{Im}[f_{i \rightarrow j}(\theta=0)]$ and rewrite the generalized optical theorem (Eq. \ref{eq:optical_theoremsym}) as:
\begin{equation}
\sigma_{sup}^{tot}=\boldsymbol{a^{\dagger}\Sigma a},
\end{equation}
where  $\boldsymbol{a}$ is a vector with the components $a_i$.

The optimization problem of finding  $\sigma_{min}^{tot}$ and $ \sigma_{max}^{tot}$ transforms to the solution of an eigenvalue equation for $\boldsymbol{\Sigma}$:
\begin{equation}
\boldsymbol{\Sigma a}=\sigma_{opt}^{tot} \boldsymbol{a},
\label{eq:opti_diag}
\end{equation}
where the optimized coefficients are the corresponding eigenvectors. The matrix $\boldsymbol{\Sigma}$ is block diagonal with respect to the symmetry of the scattering. For example, a superposition of states with different projections $M_{int}$ of the total internal angular momentum does not display interference since the imaginary part of the inelastic scattering amplitude between these states is zero. Therefore, they occupy different blocks of the matrix $\boldsymbol{\Sigma}$. The resulting eigenvectors only contain the states with the same value of $M_{int}$ and correspond to entangled superpositions thereof.

The best system controllability is obtained if the lowest eigenvalue of $\boldsymbol{\Sigma}$ is equal to zero; i.e. if the determinant of $\boldsymbol{\Sigma}$ is null. As the matrix $\boldsymbol{\Sigma}$ is semi-infinite ($\textrm{det}(\boldsymbol{\Sigma}) \geq 0$), we can define a generalized control index $R_c$:
\begin{equation}
R_c=\sqrt[N]{\frac{\textrm{det}(\boldsymbol{\Sigma})_-}{\textrm{det}(\boldsymbol{\Sigma})_+}},
\label{eq:cont_index_N}
\end{equation}
where $\textrm{det}(\boldsymbol{\Sigma})_\pm$ are the sum of positive and negative terms of the determinant, respectively. For example, in the two-states case, the determinant is equal to $\textrm{Im}[f_{1 \rightarrow 1}]\textrm{Im}[f_{2 \rightarrow 2}]-(\textrm{Im}[f_{1 \rightarrow 2}])^2$, and we recover expression (\ref{eq:Nc_2}). For the three states case, for example, the matrix $\boldsymbol{\Sigma}$ is then defined as:
\begin{equation}
\boldsymbol{\Sigma}=\frac{4\pi}{k}\begin{pmatrix}\textrm{Im}[f_{1 \rightarrow 1}]& \textrm{Im}[f_{1 \rightarrow 2}] & \textrm{Im}[f_{1 \rightarrow 3}] \\ \textrm{Im}[f_{1 \rightarrow 2}] & \textrm{Im}[f_{2 \rightarrow 2}] & \textrm{Im}[f_{2 \rightarrow 3}] \\ \textrm{Im}[f_{1 \rightarrow 3}] & \textrm{Im}[f_{2 \rightarrow 3}] & \textrm{Im}[f_{3 \rightarrow 3}]
\end{pmatrix}.
\end{equation}
The determinant of this matrix is given by:
\begin{equation}
\begin{split}
&\textrm{det}(\boldsymbol{\Sigma})=\frac{4\pi}{k}\Bigg(\textrm{Im}[f_{1 \rightarrow 1}] \Big(\textrm{Im}[f_{2 \rightarrow 2}]\textrm{Im}[f_{3 \rightarrow 3}]-\textrm{Im}[f_{2 \rightarrow 3}]^2\Big)\\-&\textrm{Im}[f_{1 \rightarrow 2}]\Big(\textrm{Im}[f_{1 \rightarrow 2}]\textrm{Im}[f_{3 \rightarrow 3}]-\textrm{Im}[f_{2 \rightarrow 3}]\textrm{Im}[f_{1 \rightarrow 3}]\Big)\\+&\textrm{Im}[f_{1 \rightarrow 3}] \Big(\textrm{Im}[f_{1 \rightarrow 2}]\textrm{Im}[f_{2 \rightarrow 3}]-\textrm{Im}[f_{2 \rightarrow 2}]\textrm{Im}[f_{1 \rightarrow 3}]\Big)\Bigg).
\end{split}
\label{eq:det_3}
\end{equation}
Using the definition (\ref{eq:cont_index_N}), the control index takes the form:
\begin{equation}
R_c=\sqrt[3]{\frac{\textrm{det}(\boldsymbol{\Sigma})_-}{\textrm{det}(\boldsymbol{\Sigma})_+}},
\label{eq:N_c_3}
\end{equation}
where:
\begin{equation}
\begin{split}
\textrm{det}(\boldsymbol{\Sigma})_-=&\textrm{Im}[f_{1 \rightarrow 1}](\textrm{Im}[f_{2 \rightarrow 3}])^2+\textrm{Im}[f_{2 \rightarrow 2}](\textrm{Im}[f_{1 \rightarrow 3}])^2\\+&\textrm{Im}[f_{3 \rightarrow 3}](\textrm{Im}[f_{1 \rightarrow 2}])^2,
\end{split}
\end{equation}
and
\begin{equation}
\begin{split}
\textrm{det}(\boldsymbol{\Sigma})_+=&\textrm{Im}[f_{1 \rightarrow 1}]\textrm{Im}[f_{2 \rightarrow 2}]\textrm{Im}[f_{3 \rightarrow 3}]\\+&2\textrm{Im}[f_{1 \rightarrow 2}]\textrm{Im}[f_{2 \rightarrow 3}]\textrm{Im}[f_{1 \rightarrow 3}],
\end{split}
\end{equation}
when $\textrm{Im}[f_{1 \rightarrow 2}]\textrm{Im}[f_{2 \rightarrow 3}]\textrm{Im}[f_{1 \rightarrow 3}]>0$. Otherwise the term $2\textrm{Im}[f_{1 \rightarrow 2}]\textrm{Im}[f_{2 \rightarrow 3}]\textrm{Im}[f_{1 \rightarrow 3}]$ must be transferred to $\textrm{det}(\boldsymbol{\Sigma})_-$. As stated previously, the value of $R_c$ can be used to interpret systems and coherent superpositions that give large control of the total ICS, as is done in the next section. Specifically, $R_c$ close to one indicates the most efficient coherent control.
\section{Systems with extensive control of the total cross section}
The theory above provides a foundation for coherent control of the total ICS. In the next section, we consider the coherent control of realistic atomic and molecular collisions at ultralow temperatures. In particular, we demonstrate the possibility of extensive coherent control of total ICS for O$_2$+O$_2$ and Rb+Rb collisions.
\subsection{O$_2$+O$_2$ scattering}
\begin{figure*}
	\centering
\includegraphics[width=1.8\columnwidth]{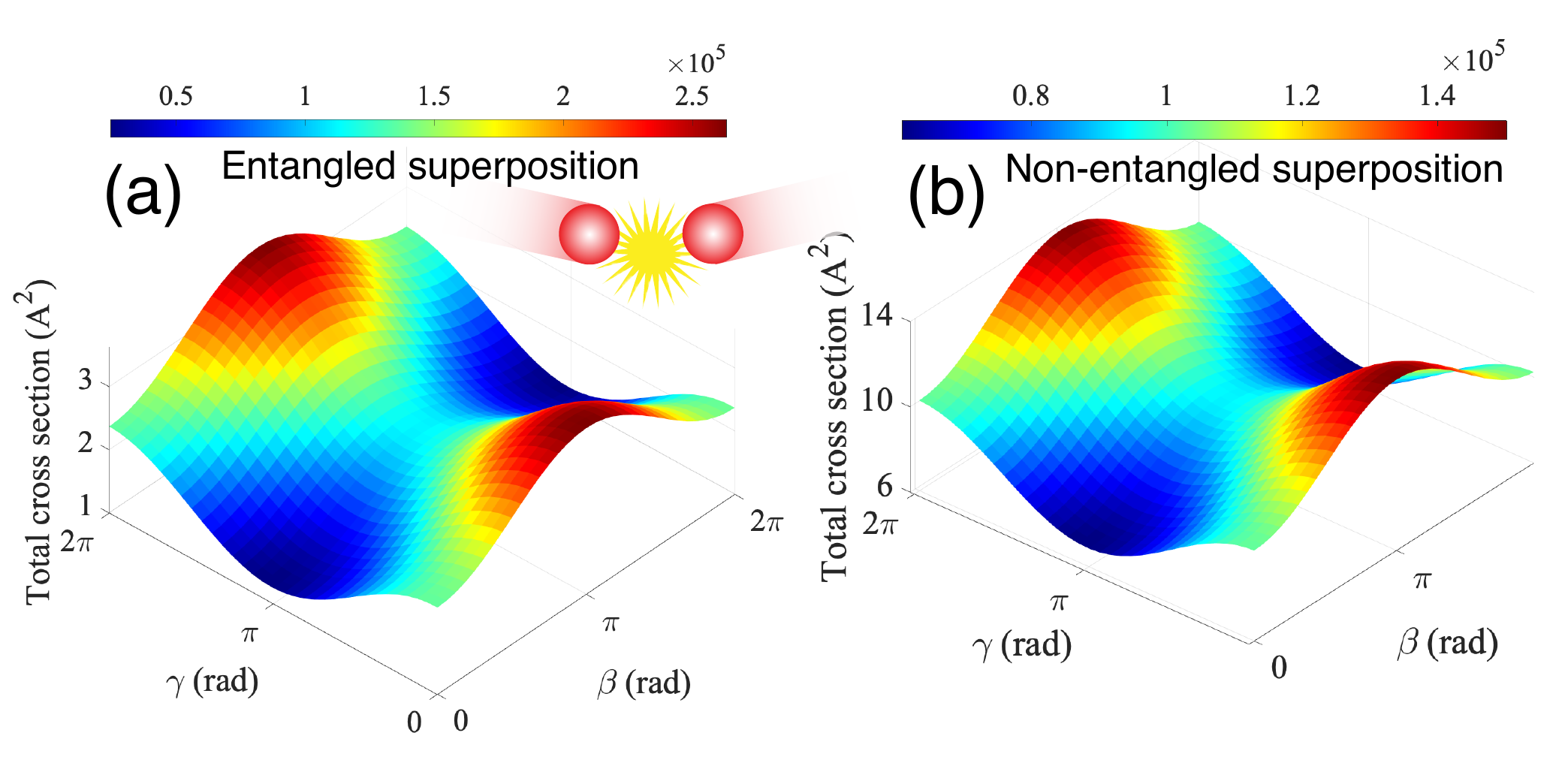}
	\caption{Coherent control of the total cross section for the ultracold $^{85}$Rb+$^{85}$Rb collisions in their lower hyperfine states $F=2$ at 50 $\mu$K from the initial superpositions (a) $\Psi_{E}$ and (b) $\Psi_{NE}$. The relative phases $\beta,\gamma$ are varied, while the relative populations $\eta=11\pi/32$, $\epsilon=3\pi/8$ are fixed for achieving the best control.}
	\label{fig:RbRb}
\end{figure*}
The first system considered is ultracold scattering of two oxygen molecules in their rovibrational ground states at 10 $\mu$K. This system has been realized experimentally in a magnetic trap at 50 mK \cite{Segev:19}, with further evaporative or sympathetic cooling projected to achieve the $\mu$K regime. Hence, it is an advantageous system for study. The oxygen molecules have spin 1, and a spin-exchange processes can occur during the scattering. The initial internal states considered here is an entangled superposition between the symmetrized states $\ket{S_A=1,M_{S_A}=-1,S_B=1,M_{S_B}=1}$ and $\ket{S_A=1,M_{S_A}=0,S_B=1,M_{S_B}=0}$:
  \begin{equation}
  \Psi_{E}(\vec{r},\xi)=\textrm{e}^{ikz} \, \left(\cos\eta\ket{1,-1,1,1}+\,\sin\eta \textrm{e}^{i\beta}\ket{1,0,1,0}\right),
  \label{eq:sup_en_O2_O2}
  \end{equation}
where $M_{S_{A/B}}$ is the projection of the electronic spin $S_{A/B}$ along the space-fixed $z$ axis. The symmetrized initial states of identical bosons are defined as: 
\begin{equation}
\begin{split}
\ket{S_1,m_1,S_2,m_2}=\frac{1}{\sqrt{2(1+\delta_{1,2})}}\Big[&\ket{S_1,m_1}_A\ket{S_2,m_2}_B\\+&\ket{S_2,m_2}_A\ket{S_1,m_1}_B \Big],
\end{split}
\end{equation} 
where $m_2\ge m_1$, and where the subscript on, for example, $\ket{S_1,m_1}_A$ denotes the states of particle A. 
\\

The scattering results (see Appendix D for computational details) show that the control index is close to 1, i.e $R_c=0.97$ ($\textrm{Im}[f_{1 \rightarrow 1}]$=158.96 a.u., $\textrm{Im}[f_{2 \rightarrow 2}]$=162.39 a.u. and $\textrm{Im}[f_{1 \rightarrow 2}]$=-156.00 a.u.). Figure \ref{fig:O2O2} (a) shows the total ICS as a function of the relative population $\eta$ and phase $\beta$ of the initial superposition (\ref{eq:sup_en_O2_O2}). The minimum value of the total ICS is seen to be 68 times smaller than the maximum value (11745 and 795427 $\AA^2$, respectively), confirming the analysis in terms of the control index. This control can also be analyzed by decomposing the total cross section into the different state-to-state cross sections. Two explanations of the large extent of control can be identified. First, in accord with the Wigner threshold law \cite{Hermsmeier2021}, only two final channels ($\ket{1,-1,1,1}$ and $\ket{1,0,1,0}$) substantially contribute to the total cross section. Second, there is complete control of the state-to-state cross sections to these two channels, as illustrated in our earlier work \cite{Devolder:21}.

\subsection{Rb+Rb scattering}
A second system of interest is the ultracold scattering of $^{85}$Rb atoms in their lower hyperfine states $F=2$ at 50 $\mu$K. This system can be readily realized experimentally in either an optical dipole trap \cite{Grimm2000,Courteille1998} or an optical tweezer \cite{Sompet2019}, allowing for precise control over internal and external atomic states. The scattering state is prepared in an entangled superposition of three symmetrized states $\ket{F_1=2,M_{F_1}=-2,F_2=2,M_{F_2}=2}$, $\ket{F_1=2,M_{F_1}=-1,F_2=2,M_{F_2}=1}$ and $\ket{F_1=2,M_{F_1}=0,F_2=2,M_{F_2}=0}$:
\begin{equation}
\begin{split}
\Psi_{E}(\vec{r},\xi)=\textrm{e}^{ikz} \Big(&\cos\eta \,\sin\epsilon\ket{2,-2,2,2}\\+&\sin\eta \, \sin\epsilon \,\textrm{e}^{i\beta}\ket{2,-1,2,1}\\+&\cos\epsilon \ \textrm{e}^{i\gamma} \ket{2,0,2,0}  \Big).
\label{eq:sup_en_Rb_Rb}
\end{split}
\end{equation}
Here, $M_F$ denotes the projection of the total angular momentum $\vec{F}=\vec{I}+\vec{S}$ along the z axis. The angles $\eta,$$\epsilon \in [0,\pi/2]$ determine the relative populations, and $\beta,\gamma \in [0,2\pi]$ the relative phases of the superposition.\\
In this case, the value of the control index (Eq. (\ref{eq:N_c_3})) is $0.83$, showing the robustness of control. We find that the minimum value of the total cross section (see Fig.\ref{fig:RbRb} (a)) is 11 times smaller than the maximal value (24,669 and 263,065 $\AA^2$ respectively). Here, the control is less effective than for the case of the scattering of oxygen molecules, but the advantage of the rubidium case is in experimental implementation. That is, ultracold rubidium atoms are widely available in optical dipole traps \cite{Grimm2000,Courteille1998} or, more recently, in optical tweezer setups \cite{Sompet2019}. Furthermore, initial steps toward the creation of entangled superposition similar to Eq.(\ref{eq:sup_en_Rb_Rb}) between two $^{85}$Rb atoms have been demonstrated in optical tweezers\cite{Sompet2019}. 
\subsection{Non-entangled superpositions}
The superpositions (\ref{eq:sup_en_O2_O2}) and (\ref{eq:sup_en_Rb_Rb}) are entangled and cannot be obtained via independent preparation of each of the collision partners. That poses an experimental challenge, since independent preparation is the easiest way to experimentally realize coherent control. To examine the effect of initial state entanglement, we consider scattering of unentangled independently prepared molecules.

For the case of O$_2$+O$_2$ collisions, the two colliding molecules are prepared in two different superpositions:
\begin{align}
\ket{\psi_A}&=N\left(\sqrt{\cos\eta} \ket{1,-1}+\sqrt{\sin\eta}\,\textrm{e}^{i\frac{\beta}{2}}\ket{1,0}\right) \\
\ket{\psi_B}&=N\left(\sqrt{\sin\eta}\,\textrm{e}^{i\frac{\beta}{2}}\ket{1,0}+\sqrt{\cos\eta} \ket{1,1}\right),
\end{align}
where $N=(\sin\eta+\cos\eta)^{-1/2}$ is a normalization factor. Here, for simplicity, we consider states of A and B with population and relative phase in both being determined by the same parameter $\eta$ and $\beta$. Different choices of control parameters in A and B are also possible.\\
The initial superposition is then obtained by symmetrizing the product $\ket{\psi_A}\ket{\psi_B}$ to give:
\begin{multline}\label{psiS2}
\Psi_{NE}(\vec{r},\xi)=N^2 \textrm{e}^{ikz} \Big[\cos\eta \ket{{1,-1,1,1}}+\sin\eta \,\textrm{e}^{i\beta}\ket{{1,0,1,0}}\\+\sqrt{\cos\eta\,\sin\eta}\,\textrm{e}^{i\frac{\beta}{2}} (\ket{{1,-1,1,0}}+\ket{{1,0,1,1}})\Big].
\end{multline}
Note that Eq. (\ref{psiS2}) is unentangled as compared to Eq. (\ref{eq:sup_en_O2_O2}). Preparation at the level of the individual molecules implies that the superposition contains a range of states with different values of $M_{int}$. Here, there are two states with $M_{int}$=0, one state with $M_{int}$=1 and one state with $M_{int}$=-1. Since these supplementary states do not interfere, the extent of the control is reduced. Also, the normalization factor, which depends on $\eta$, has an impact on the control. However, with $\eta=\pi/4$ fixed (see Fig. \ref{fig:O2O2}(d)), coherent control by varying the relative phase is still extensive, with a factor of 41 between the maximal and minimal values. Therefore, even with a non-entangled superposition, it is possible to substantially control the total cross section for O$_2$+O$_2$ scattering.

As an example in the Rb+Rb case, the two colliding atoms are prepared in two different three-state superpositions:
\begin{equation}
\begin{split}
\ket{\psi_A}=N_3\Big(&\sqrt{\sin\eta \,\sin\epsilon} \ket{2,-2}\\+&\sqrt{\cos\eta\,\sin\epsilon}\,\textrm{e}^{i\frac{\beta}{2}}\ket{2,-1}\\+&\sqrt{\cos\epsilon}\,\textrm{e}^{i\frac{\gamma}{2}}\ket{2,0}\Big),
\end{split}
\end{equation}
\begin{equation}
\begin{split}
\ket{\psi_B}=N_3\Big(&\sqrt{\sin\eta \,\sin\epsilon} \ket{2,2}\\+&\sqrt{\cos\eta \,\sin\epsilon}\,\textrm{e}^{i\frac{\beta}{2}}\ket{2,1}\\+&\sqrt{\cos\epsilon}\,\textrm{e}^{i\frac{\gamma}{2}}\ket{2,0}\Big),
\end{split}
\end{equation}
where $N_3=(\sin\eta \,\sin\epsilon+\cos\eta\,\sin\epsilon+\cos\epsilon)^{-1/2}$. \\
As in the oxygen case, the initial non-entangled superposition is obtained by symmetrizing the product $\ket{\psi_A}\ket{\psi_B}$:
\begin{multline}\label{Rb_nen}
\Psi_{NE}(\vec{r},\xi)=N_3^2 \, \textrm{e}^{ikz} \Big[\cos\eta \, \sin\epsilon\ket{2,-2,2,2}\\+\sin\eta \,\sin\epsilon \, \textrm{e}^{i\beta}\ket{2,-1,2,1}+\cos\epsilon \, \textrm{e}^{i\gamma} \ket{2,0,2,0} \\ +\sqrt{\sin\eta \,\cos\eta}\,\sin\epsilon \,\textrm{e}^{i\frac{\beta}{2}} (\ket{2,-2,2,1}+\ket{2,-1,2,2})\\
+\sqrt{\sin\eta\,\sin\epsilon \,\cos\epsilon}\,\textrm{e}^{i\frac{\gamma}{2}}(\ket{2,-2,2,0}+\ket{2,0,2,2}) \\
+\sqrt{\cos\eta\,\sin\epsilon\,\cos\epsilon}\,\textrm{e}^{i\frac{\beta+\gamma}{2}}(\ket{2,-1,2,0}+\ket{2,0,2,1})\Big].
\end{multline}
In this case, there are four groups of states: one group of three states with $M_{int}=0$, two groups of two states with $M_{int}=\pm 1$ and two groups of one state with $M_{int}=\pm 2$. The total cross section is the sum of the ICS correspondingly to each group. Coherent control by varying the relative phases is shown in Figure \ref{fig:RbRb} (b). Here, the ratio max/min is significantly reduced,relative to the entangled case, to a factor of two (163,911 and 74,512 $\AA^2$ respectively). With the non-entangled superposition, the quasi-vanishing of the total cross section is lost. This difference between the oxygen and rubidium cases arises from the number of uncontrollable "satellite" terms \cite{Brumer_book}, prevalent in Eq. (\ref{Rb_nen}). Nevertheless, if the cross section can be accurately measured, the predicted control is sufficiently large to be experimentally measurable.
\section{Conclusion}
We derived the generalized multichannel optical theorem for the scattering of initial coherent superpositions, a fundamental contribution to scattering theory, introducing new interference contributions. As an example, it was then used to address three issues. The first was to determine how the generalized optical theorem reflects contributions from an initial superposition of internal states. As in the standard form of the optical theorem, the forward scattering plays an essential role. However, the generalized theorem shows that inelastic scattering between the states involved in the initial superpositions is  crucial in the optical theorem and here in applications to the coherent control of the total integral cross section.

The second issue concerned  optical theorem insights into the coherent control of the total scattering cross section. The maximal and the minimal values of the total cross sections were found to be directly related to the elastic and inelastic forward scattering amplitudes. If these quantities fulfill the Schwartz equality, the minimal value vanishes and complete control is possible. Furthermore, the Schwartz equality allowed us to define a control index through which the extent of control can be understood. These statements were generalized to include the initial superpositions of $N_{sup}$ internal states of the collision partners. 

Finally, the third issue concerned identifying cold atomic and molecular collision systems with a large extent coherent control to motivate experimental demonstrations of coherent control of scattering. Two interesting cases were examined: O$_2$+O$_2$ scattering in their rovibrational ground state, and the scattering of rubidium atoms $^{85}$Rb. To examine experimental requirements on the initial state preparation, we considered both entangled and non-entangled superpositions, where the latter could be created by preparation of the individual molecules. Although the entangled states display greater control, the non-entangled initial superpositions show sufficient control to be both significant and experimentally observable. Ultracold O$_2$+O$_2$ scattering was found to allow better control, while ultracold $^{85}$Rb+$^{85}$Rb collisions could be easier to probe experimentally, given widespread availability of ultracold Rb atoms in either optical dipole traps \cite{Grimm2000,Courteille1998} or optical tweezers \cite{Sompet2019}.  {These two cases demonstrate that a large degree of control can be obtained for multi-channel scattering, and} motivate the first experimental demonstration of the quantum interference based control of the total cross section in ultracold atomic and molecular collisions.
\section*{Acknowledgements} 
This work was supported by the U.S. Air Force Office of Scientific Research (AFOSR) under Contract No.
FA9550-19-1-0312. SciNet computational facilities are gratefully acknowledged.
\section*{Appendix A: Calculation of the incoming flux}
The incoming current density for the superposition state takes the form:
\begin{equation}
\begin{split}
\vec{j}_{in}(\vec{r})=\Bigg[\int \textrm{d}\xi \frac{\hbar}{2i\mu}\left(\sum_{i=1}^{N_{sup}}a_i\ket{i}\textrm{e}^{ikz}\right)^*\\\vec{\nabla}\left(\sum_{i'=1}^{N_{sup}}a_{i'}\ket{i'}\textrm{e}^{ikz}\right)\Bigg]+\textrm{c.c}.
\end{split}
\end{equation}
Using the orthogonality of the channel basis states $\braket{i|i'}=\delta_{i,i'}$ and  $\sum_{i=1}^{N_{sup}}|a_i|^2=1$, one obtains: 
\begin{equation}
\vec{j}_{in}(\vec{r})=\frac{\hbar k}{\mu}\hat{e}_z,
\end{equation}
where $\hat{e}_z$ is the unit vector along $\vec{z}$. \\
The integral on the closed surface gives:
\begin{equation}
I_{in}=\lim_{r\rightarrow\infty}r^2 \frac{\hbar k}{\mu}\int\hat{e}_z\cdot\hat{e}_r \textrm{d}\Omega,
\end{equation}
\begin{equation}
I_{in}=\lim_{r\rightarrow\infty}r^2 \frac{\hbar k}{\mu} 2\pi\int_0^\pi \,\sin\theta \, \cos\theta \textrm{d}\theta.
\end{equation}
Since $\int_0^\pi \,\sin\theta \, \cos\theta \textrm{d}\theta=0$, the incoming contribution vanishes:
\begin{equation}
I_{in}=0.
\end{equation}
\section*{Appendix B: Calculation of the outgoing flux}
The outgoing current density is: 
\begin{equation}
\begin{split}
\vec{j}_{out}(\vec{r})=\Bigg[\int \textrm{d}\xi\frac{\hbar}{2i\mu}\left(\sum_{j } f_{sup \rightarrow j}(\theta,\phi) \frac{\textrm{e}^{ik_jr}}{r}\ket{j}\right)^*\\\vec{\nabla}\left(\sum_{j' } f_{sup \rightarrow j'}(\theta,\phi) \frac{\textrm{e}^{ik_{j'}r}}{r}\ket{j'}\right)\Bigg]+\textrm{c.c}.
\label{eq:out_curr_dens}
\end{split}
\end{equation}
Using the orthogonality relation, equation (\ref{eq:out_curr_dens}) becomes:
\begin{equation}
\vec{j}_{out}(\vec{r})=\frac{\hbar}{\mu}\sum_j \frac{k_j}{r^2}|f_{sup \rightarrow j}(\theta,\phi)|^2 \hat{e}_r +O\left(\frac{1}{r^3}\right).
\end{equation}
Note that the angular components of the gradient operator (included in $O\left(\frac{1}{r^3}\right)$) have been neglected. Indeed, when the limit of $r$ to infinity is taken, those angular components become negligible in comparison to the radial component. 

The integrated quantity takes the form:
\begin{equation}
I_{out}=\frac{\hbar}{\mu} \sum_{j} k_j\int |f_{sup \rightarrow j}(\theta,\phi)|^2 \textrm{d}\Omega.
\end{equation}
Using the definitions of the differential cross section , $\frac{d\sigma_{sup}}{d\Omega}=\frac{k_j}{k}|f_{sup \rightarrow j'}(\theta,\phi)|^2$,  and of the total ICS, $\sigma_{sup}^{tot}=\sum_{j}\int \frac{d\sigma_{sup}}{d\Omega} d\Omega$, $I_{out}$ becomes:
\begin{equation}
I_{out}=\frac{\hbar k}{\mu}\sigma_{sup}^{tot}.
\end{equation}
\section*{Appendix C: Calculation of the interfering flux}

The interference current density takes the form: 
\begin{equation}
\begin{split}
\vec{j}_{int}(\vec{r})=\Bigg[&\int \textrm{d}\xi\frac{\hbar}{2i\mu}\left(\sum_{i=1}^{N_{sup}}a_i\ket{i}\textrm{e}^{ikz}\right)^*\\&\vec{\nabla}\left(\sum_{j} f_{sup \rightarrow j}(\theta,\phi) \frac{\textrm{e}^{ik_jr}}{r}\ket{j}\right) \\
+&\frac{\hbar}{2i\mu}\left(\sum_{j} f_{sup \rightarrow j}(\theta,\phi) \frac{\textrm{e}^{ik_jr}}{r}\ket{j}\right)^*\\&\vec{\nabla}\left(\sum_{i=1}^{N_{sup}}a_i\ket{i}\textrm{e}^{ikz}\right)\Bigg]+\textrm{c.c}.
\end{split}
\label{eq:interference_1}
\end{equation}
After the applications of the gradient and the orthogonality relation, the interference current density becomes:
\begin{equation}
\begin{split}
\vec{j}_{int}(\vec{r})=\Bigg[\frac{\hbar}{2\mu} \sum_{i}^{N_{sup}}a_i^*f_{sup \rightarrow i}(\theta,\phi)\textrm{e}^{ik r(1-\cos\theta)}\\\left[\frac{k}{r}(1+\cos\theta)+\frac{i}{r^2}\right]\hat{e}_r\Bigg]+\textrm{c.c}.
\end{split}
\end{equation}
As for the outgoing contribution, only the radial component of the gradient is considered.
The integral on the closed surface is equal to:
\begin{equation}
\begin{split}
I_{int}=\Bigg[\frac{\hbar}{2\mu} \sum_{i}^{N_{sup}}a_i^*\lim_{r\rightarrow\infty}\int_0^{2\pi}d\phi\int_0^\pi \textrm{d}\theta \,\sin\theta f_{sup \rightarrow i}(\theta,\phi)\\ \textrm{e}^{ikr(1-\cos\theta)}\left[r k(1+\cos\theta)+i\right]\Bigg]+\textrm{c.c}.
\end{split}
\end{equation}
Due to the uniform convergence of the limit, the order of the integral on $\phi$ and the limit on $r$ can be interchanged. We focus in the $\theta$ integral first. Then, two limits must be calculated:
\begin{equation}
\lim_{r\rightarrow\infty}r k\int_0^\pi \textrm{d}\theta \,\sin\theta f_{sup \rightarrow i}(\theta,\phi)\textrm{e}^{ikr(1-\cos\theta)}(1+\cos\theta),
\label{eq:limit_1}
\end{equation}
\begin{equation}
\lim_{r\rightarrow\infty}i\int_0^\pi \textrm{d}\theta \,\sin\theta f_{sup \rightarrow i}(\theta,\phi)\textrm{e}^{ikr(1-\cos\theta)}.
\label{eq:limit_2}
\end{equation}
We consider the second limit(\ref{eq:limit_2}) and we make the changes of variables $x=1-\cos\theta$ and $\kappa=kr$. We also, for notational clarity, suppress the $\phi$ dependence of $f_{sup \rightarrow i}(\theta,\phi)$ until Eq.(\ref{eq: I_int}). Eq. (\ref{eq:limit_2}) then becomes:
\begin{equation}
\begin{split}
\lim_{r\rightarrow\infty}\int_0^\pi \textrm{d}\theta \,\sin\theta f(\theta)\textrm{e}^{ikr(1-\cos\theta)}=\\ \lim_{\kappa\rightarrow\infty}\int_0^2 \textrm{d}x f(x)\textrm{e}^{i\kappa x}.
\end{split}
\end{equation}
By the Riemann-Lebesgue lemma, this limit vanishes:
\begin{equation}
 \lim_{\kappa\rightarrow\infty}\int_0^2 \textrm{d}x f(x)\textrm{e}^{i\kappa x}=0.
\end{equation}
Now, we focus on the limit (\ref{eq:limit_1}) and make the same change of variables:
\begin{equation}
\begin{split}
\lim_{r\rightarrow\infty}kr\int_0^\pi \textrm{d}\theta \, \sin\theta f(\theta)\textrm{e}^{ikr(1-\cos\theta)}(1+\cos\theta)=\\ \lim_{\kappa\rightarrow\infty}\kappa\int_0^2 \textrm{d}x f(x)\textrm{e}^{i\kappa x}(2-x).
\end{split}
\end{equation}
After integration by parts, we obtain:
\begin{equation}
\begin{split}
\lim_{\kappa\rightarrow\infty}\kappa\int_0^2 \textrm{d}x f(x)\textrm{e}^{i\kappa x}(2-x)=\\\frac{1}{i}\Bigg(\lim_{\kappa\rightarrow\infty}[f(x)(2-x)\textrm{e}^{i\kappa x}]_0^2\\-\lim_{\kappa\rightarrow\infty}\int_0^2 (f(x)(2-x))'\textrm{e}^{i\kappa x}\textrm{d}x\Bigg).
\end{split}
\end{equation}
Using the Riemann-Lebesgue lemma, the second term evaluates to zero:
\begin{equation}
\begin{split}
\lim_{\kappa\rightarrow\infty}\kappa\int_0^2 \textrm{d}x f(x)\textrm{e}^{i\kappa x}(2-x)=\\\frac{1}{i}\left(\lim_{\kappa\rightarrow\infty}[f(x)(2-x)\textrm{e}^{i\kappa x}]_0^2\right).
\end{split}
\end{equation}
The first term gives:
\begin{equation}
\lim_{\kappa\rightarrow\infty}\kappa\int_0^2 \textrm{d}x f(x)\textrm{e}^{i\kappa x}(2-x)=\frac{-2f(x=0)}{i},
\end{equation}
\begin{equation}
\lim_{\kappa\rightarrow\infty}\kappa\int_0^2 \textrm{d}x f(x)\textrm{e}^{i\kappa x}(2-x)=2if(x=0).
\end{equation}
$x=0$ corresponds to $\theta=0$. Therefore, one obtains:
\begin{equation}
\begin{split}
&\lim_{r\rightarrow\infty}kr\int_0^\pi \textrm{d}\theta \, \sin\theta f(\theta)\textrm{e}^{ikr(1-\cos\theta)}(1+\cos\theta)=\\& 2if(\theta=0).
\end{split}
\end{equation}
Then, $I_{int}$ becomes, where we restore the $\phi$ dependence in $f$ and consider the $\phi$ integral:
\begin{equation}
I_{int}=\frac{i\hbar}{\mu}\sum_{i}^{N_{sup}}a_i^*\int_{0}^{2\pi} \textrm{d}\phi f_{sup \rightarrow i}(\theta=0,\phi)+\textrm{c.c}.
\label{eq: I_int}
\end{equation}
The forward scattering amplitudes from the superposition $f_{sup \rightarrow i}(\theta=0,\phi)$ is then expanded in state-to-state scattering amplitudes (see Eq. (\ref{eq:scatt_amp_sup})):
\begin{equation}
I_{int}=\frac{i\hbar}{\mu}\sum_{i,i'=1}^{N_{sup}}a_i^*a_{i'} \int_{0}^{2\pi} \textrm{d}\phi f_{i' \rightarrow i}(\theta=0,\phi)+\textrm{c.c}.
\end{equation}
The forward scattering amplitude does not depend on $\phi$, so that the integral on $\phi$ simply gives:

\begin{equation}
I_{int}=\frac{2\pi i\hbar}{\mu}\sum_{i,i'=1}^{N_{sup}}a_i^*a_{i'} f_{i' \rightarrow i}(\theta=0)+\textrm{c.c},
\end{equation}
\begin{equation}
I_{int}=\frac{-4\pi\hbar}{\mu} \sum_{i,i'=1}^{N_{sup}} \textrm{Im}[a_i^*a_{i'}f_{i' \rightarrow i}(\theta=0)].
\end{equation}
\section*{Appendix D: Details of the numerical calculations}
\subsection*{O$_2$+O$_2$ scattering}
We perform quantum scattering calculations on O$_2$~+~O$_2$ collisions using a coupled-channel  (CC)  methodology \cite{Tscherbul2008} based on the expansion of the scattering wavefunction in an 
uncoupled symmetrized space-fixed  basis set composed of direct products of molecular rotational and spin basis functions and the orbital angular momentum eigenstates. Most of the computational details are essentially the same as reported in the previous work of one of the authors \cite{Tscherbul2008}. The CC basis set was composed of  three rotational states ($N=0{-}4$) and 6 partial waves ($\ell=0{-}10$) at 10 $\mu$K. The hyperfine structure of $^{17}$O$_2$ was neglected to make the calculations computationally feasible.
The CC equations were integrated on the radial grid  from  $R_\text{min}=4.0\,a_0$ to $R_\text{max}=150\, a_0$   with a grid step of $0.04\,a_0$.
The T-matrix elements are obtained from this CC results and are used to calculate the forward scattering amplitudes:
\begin{equation}
f_{i\rightarrow j}(\theta=0)=\frac{1}{2k} \sum_{\ell}\sum_{\ell'} i^{\ell-\ell'+1}\sqrt{(2\ell+1)(2\ell'+1)} T_{i\ell 0 \rightarrow j\ell'0}.
\end{equation}
Finally, the total cross section is calculated using the generalized optical theorem (Eq. \ref{eq:optical_theoremsym}).
\subsection*{Rb+Rb scattering}
Quantum scattering calculations of ultracold $^{85}$Rb+ $^{87}$Rb collisions were performed following the same methodology as in the previous works \cite{Hermsmeier2021,Li2007,Li:08}. The CC equations were integrated on the radial grid  from  $R_\text{min}=2.0\,a_0$ to $R_\text{max}=300\, a_0$   with a grid step of $0.005\,a_0$. The calculation of the forward scattering amplitude and of the total cross section is the same as that for O$_2$-O$_2$ scattering. 
\bibliography{optical_theorem}
\end{document}